\documentclass[12pt]{iopart}
\usepackage{graphicx}
\newfont{\Bb}{msbm10 scaled\magstep1}

\begin{document}
\title[Integrability of one degree of freedom symplectic maps]
{Integrability of one degree of freedom symplectic maps with polar singularities}
\date{}
\author{Minoru Ogawa \\
Department of Physics, Nagoya University, \\ Nagoya 464-8602 JAPAN \\
electric address: mogawa@r.phys.nagoya-u.ac.jp}

\begin{abstract}
In this paper, we treat symplectic difference equations with one degree of freedom.
For such cases, we resolve the relation between that the dynamics on the two dimensional phase
space is reduced to on one dimensional level sets by a conserved
quantity and that the dynamics is integrable, under some assumptions. The process which we
introduce is related to interval exchange transformations.
\end{abstract}
\section{Introduction}
Exploring the behaviour of systems is an old matter of interest for researchers on dynamical systems
\cite{L}\cite{P}. It is varies from integrable to chaos; see for example \cite{AA}.
\par
The question whether a system is integrable or not is one of the most interested
matters \cite{A1}.
For autonomous ordinary differential equations, integrability means that the equation is solvable
by quadratures. Complete integrability is one of the most simple and useful
integrabilities for flows. It is based on the fact that we can solve any one dimensional flows.
Note that, in the case of Hamiltonian flows, Liouville's integrability, instead of complete
integrability, is frequently used because of its topological aspect.
\par
Here we recall the concept of complete integrability. For $n$th order autonomous ordinary
differential equations, we
say that a system is completely integrable if there exist $n-1$ conserved quantities and
they are functionally independent. Functionally independent conserved quantities are called
first integrals, since their values arise as integral constants in the general solution.
\par
For autonomous ordinary difference equations, however, complete integrability can not be worked on
since we can not solve one dimensional maps in general. The map $T:[0,1]\to [0,1]$ defined as
$Tx\equiv rx(1-x)$, $0<r\leq 4$ is such a typical example. Therefore integrability for autonomous
ordinary difference equations, or maps, is more complicated than flows. Note that even in the
case of one dimensional maps with invertibility, the dynamics may not be solvable.
Indeed, many types of interval exchange transformations with generic parameters seem not to be
solvable since they have weak mixing property \cite{V1}.
\par
While for partial difference equations, researches of intgegrability for the systems which
correspond to soliton equations were developed in these two decades by many researchers;
see for example \cite{Hirota}.
\par
In this paper, we give a sufficient condition that a symplectic difference equation with one
degree of freedom, which are naturally equipped with invertibility,
is solvable by quadratures. More precisely, for a symplectic difference equation
\[
x_{n+1}+x_{n-1}=f(x_n),\quad f(x)\mbox{ is continuous or rational,}
\]
with one degree of freedom,
if the system has a conserved quantity $I(x_n,x_{n+1})$ and there exists an invariant finite
Borel measure $\mu_C$ on each level set $I=C$, then the system is solvable by quadratures.
This integrability can be regarded as an analogue of complete integrability for flows.
\par
For difference equations with singularities, singularity confinement test \cite{GRP}
is available for checking whether the system has sufficiently many conserved quantities.
Note that singularity confinement test gives us neither sufficient nor necessary condition to have
conserved quantities; see \cite{HV} for non-sufficiency and see the example presented in next
section for non-necessarity,
nevertheless, many difference equations with conserved quantities pass this test.
\par
In the cases of birational maps, algebraic entropy \cite{BV} is useful to judge
whether a map has conserved quantities. This method says that if algebraic entropy $E=0$, the
system is integrable. Note that this method is related to Arnold's complexity \cite{A2}.
\par
Both in singularity confinement test and in algebraic entropy method, the discussion about the
solvability by quadratures is lacked. In this paper, a sufficient condition that symplectic
difference equations can be solved by quadratures is presented through a geometrical approach.
That is, we shed a light to the geometrical aspects of integrability of symplectic maps.
\par
In the rest of this paper, we use the word ``integrability" as the meaning of that the system is
solvable by quadratures both for maps and for flows.
\par
The rest of this paper is organized as follows. In section 2, the concept of ``onset linearization"
is proposed through a typical example. In section 3, the construction of invariant measure on
each level set is presented, and then, we can see
the connection between the dynamics on each level set and interval exchange transformations under
some assumptions. In section 4, the main theorems are stated. An example whose integrability can be
proven by the main theorems is presented in appendix.

\section{``onset linearization"}
Linearizability is one of the most simple integrabilities both for flows and for maps.
Consider $k$th order autonomous difference equations
\begin{equation}
x_{n+k}=f(x_n,\dots ,x_{n+k-1})\quad x\in\mbox{\Bb R}\mbox{ (, or }\in\mbox{\Bb C}\mbox{)}.
\label{general}
\end{equation}
If there exists a transformation $g:\mbox{\Bb R}\to\mbox{\Bb R}$(,or $\mbox{\Bb C}\to \mbox{\Bb C}$)
such that the difference equation (\ref{general}) is rewritten as a linear difference equation
\begin{equation}
y_{n+k}=f^{\prime }(y_n,\dots ,y_{n+k-1}),\quad
f^{\prime }\mbox{ is a linear function w.r.t. each } f_{n+i},
\end{equation}
for new variable $y_n\equiv g(x_n)$,
then we say that the original difference equation (\ref{general}) is linearizable.
The concept of linearizability works on flows, too.
If a map, or a flow, is linearizable, its integrability is obvious, so linearizability is
a sufficient condition that a system is integrable. But it isn't a necessary condition,
and moreover, there exist many examples which is integrable but is not linearizable.
Such a typical eample is the dynamics of pendulum moving on a perpendicular plane.
This system has two fixed points on its cylindrical phase space,
one is elliptic and the other is hyperbolic. Around the elliptic fixed point,
the orbits don't have isochronism property which linear oscillator has.
Isochronism property is conserved under variable transformations, so this system isn't
linearizable.
\subsection{typical example}
The same situation is occurred for maps. Consider the difference equation
\begin{equation}
x_{n+1}+x_{n-1}=\frac{1}{x_n},\quad x_n\in\mbox{\Bb R}.
\label{linearizable}
\end{equation}
This system has a conserved quantity $I=x_nx_{n+1}(1-x_nx_{n+1})$ on its phase space
$P=\{ (x_n,x_{n+1})\ |\ x_n, x_{n+1}\in\mbox{\Bb R}\} $, and moreover, this system has another
conserved quantity $I^{\prime }=x_{2n}x_{2n+1}$ for sub dynamics
$T^2:\mbox{\Bb R}^2\to\mbox{\Bb R}^2$. Here $T:(x_n,x_{n+1})\mapsto (x_{n+1},x_{n+2})$ is the map
defined by (\ref{linearizable}) and $T^2$ means $T\circ T$.
The difference equation (\ref{linearizable}) can be partially linearizable by $I^{\prime }$ as
follows;
\begin{equation}
\left\{
\begin{array}{l}
x_{2n+1}+x_{2n-1}=\frac{1}{I^{\prime }}x_{2n+1} \\
x_{2n+2}+x_{2n}=\frac{1}{I^{\prime }}x_{2n},
\end{array}
\right.
\label{linearize}
\end{equation}
where the value of $I^{\prime }$ is invariant under the dynamics of the partially linearized
difference equations (\ref{linearize}). The value of $I^{\prime }$ varies as initial value varies,
so the linearized equations (\ref{linearize}) depends on the level set on which initial value is.
Let us call the partially linearizablity ``{\it onset linearizability}".
\par
For such initial values $(x_0,x_1)$ as $I^{\prime }(x_0, x_1)=0\mbox{ or }1$, the system
(\ref{linearize}) of difference equations breaks down. This breaking down corresponds to
the breaking down of (\ref{linearizable}) derived from $1/x_n$.
\par
In this system the singularity $(x_0,x_1)=(\mbox{any},0)$ can not be confined for any step.
In fact, for initial value $(x,\epsilon )$, we can calculate $x_{2n}=(\epsilon^{-1}x^{-1}-1)^nx$
by the second equation of (\ref{linearize}).
\subsection{definition}
To describe the general definition of ``onset linearizability", we return to the $k$th order
difference equation (\ref{general}).
Suppose that
(\ref{general}) has $l$ functionally independent conserved quantities $I_j$, $j=1,\dots ,l$.
Then its level sets $M(x_0,\dots ,x_{k-1})$ on which a given initial value $(x_0,\dots ,x_{k-1})$
exists are labeled by the set $\{ C_j\}_{j=1,\dots ,l}$ of
$l$ values $C_j\equiv I_j(x_0,\dots ,x_{k-1})$. If there exists a variable
transformation $g(x_n)\equiv y_n$ and the difference equation (\ref{general}) is deformed
to a linear difference equation
\[
y_{n+k}=f^{\prime }(y_n,\dots ,y_{n+k-1};I_1,\dots ,I_l),
\]
where $f^{\prime }$ is a linear function with respect to each $y_{n+j}$, then we say that the
difference equation (\ref{general}) is directly onset linearizable.
\par
And, in the same assumptions, if there exists a sub dynamics
\begin{equation}
(x_{n+i},x_{n+i+1},\dots ,x_{n+i+k})=\bar{f} (x_n,x_{n+1},\dots ,x_{n+k-1}),\quad i>1,
\label{indirectly}
\end{equation}
of (\ref{general}) and it is onset linearizable, then we say that the difference equation
(\ref{general}) is indirectly onset linearizable. Here $\bar{f}$ is a collection of functions,
$
\bar{f}=(\bar{f}_0,\dots ,\bar{f}_{k-1}).
$
If $i\leq k$, the sub dynamics (\ref{indirectly}) is equivalent to (\ref{general}).
While if $i>k$, informations of some sequences are lost. For example, if $k=2$ and $i=3$,
informations of $\{ x_{3n}\} $ are lost for the pair of initial value $(x_1,x_2)$.
But these lost sequences can be given by finite iterations, namely, $x_{3n}$ is immediately given
from $(x_{3n-2},x_{3n-1})$ by (\ref{general}). So, indirectly onset linearizability asserts
that the original dynamics is integrable.
\par
In the case of (\ref{linearizable}), the difference equation is indirectly onset linearizable with
$k=2$ and $i=2$.

\section{invariant measure on level set and interval exchange}
Consider one degree of freedom symplectic difference equation
\begin{equation}
x_{n+1}+x_{n-1}=f(x_n),\quad x_n\in\mbox{\Bb R}.
\label{symplectic1}
\end{equation}
This difference equation is equivalent to the system of difference equations
\begin{equation}
\left\{
\begin{array}{l}
x_{n+1}=y_n \\
y_{n+1}=-x_n+f(y_n),
\end{array}
\right.
\quad x_n,y_n\in\mbox{\Bb R}.
\label{symplectic2}
\end{equation}
In this and next sections, we use the notation (\ref{symplectic2}) instead of (\ref{symplectic1})
because of the easiness of description.
\par
Difference equations of this type have invertibility and symplectic property at regular
points on their phase spaces. Therefore, the dynamics preserves phase volume, namely, the measure
$\mu =\mbox{d}x\mbox{d}y$ is invariant.
\par
If (\ref{symplectic2}) has a conserved quantity $I(x,y)$, then there
exists the restricted invariant measure
\begin{equation}
\mu |_{I=C}(x,y)=\frac{\mbox{d}s}{(I_x^2+I_y^2)^{1/2}}=\frac{\mbox{d}s}{|\mbox{grad}I|},
\label{InvMeas}
\end{equation}
on each level set $I=C$ because of its symplectic property. 
Here $I_x$ and $I_y$ mean derivatives of $I$ with respect to $x$ and $y$
respectively, and $s$ is a variable defined as the arc-length of the curve $I=C$.

\section{main theorems}
In the same assumptions as previous section, we can conclude that, for almost every level set
$I(x,y)=C$, if the restricted invariant measure (\ref{InvMeas}) is finite, then the
restricted dynamics $T|_{I=C}$ of (\ref{symplectic2}) is isomorphic (mod 0) to an interval
exchange transfomation.
Here ``isomorphic (mod 0)" measns ``measure theoretically isomorphic".
Furthermore, if the function $f$ of (\ref{symplectic2}) is continuous or rational,
the interval exchange transformations are integrable.
\par
As the preparation to state the main theorem, we recall the notion of interval exchange
transformations \cite{V1}. Interval exchange transformations is a class of piecewise linear maps
on unit (or a finite) interval. For a piecewise linear map, if the map
is one-to-one, onto and every inclination is equal to $+1$, then we say that the piecewise linear
map is an {\it interval exchange transformation}, or simply, an {\it interval exchange}.
\par
Every interval exchange has the invariant measure $\mbox{d}x$ and orientability.
Note that, for any one dimensional orientable and bijective maps,
if a map has a finite Borel invariant measure and the map is a.e.- continuous with respect to
the measure, then there exists an interval exchange which is isomorphic (mod 0) to the map.
\par
Discrete rotations on unit interval, which are integrable, are the typical examples of interval
exchanges. Note that, to the best of the author's knowledge, it is not known whether there are
non-integrable interval exchanges, however there exist interval exchanges with weak mixing property
\cite{V1}, which seem to be non-integrable.
\par
For the symplectic difference equations (\ref{symplectic2}), if there exists a first integral
$I(x,y)$ and the restricted invariant measure (\ref{InvMeas}) on a generic level set $I=C$ is
finite, then there exists an one-to-one, onto map $\varphi $ acting on unit interval $[0,1)$
and preserving the Lebesgue measure $\nu =\mbox{d}x$, and $T|_{I=C}$ is isomorphic (mod 0)
to $\varphi $.
\par
Here a generic level set means the level set on which the almost every point is not the first type
singular point and every point is not the second type singular point.
Here the first type singular point is the singular point derived from $f$'s singularities,
where the absence of this type of singularity for a.e.- point guarantees one-to-one, onto
properties
of $T|_{I=C}$ for a.e.- point. The second one is the singular point on which the absolute value
$|\mbox{grad} I|$ vanishes, where the absence of this type of singularity, together with the
absence of the other, makes $T|_{I=C}$ to be orientable and a.e.- one-to-one and onto.
\par
Note that the orientability of $T_{I=C}$ of (\ref{symplectic2}) is directly derived from
symplectic property of (\ref{symplectic2}),
that is, we choose the orientation (cotangent vector) $\sigma (s)$ on $I=C$ by,
for example, 
\[
\left\{
\begin{array}{ll}
\sigma (x,y)\equiv 1, & \quad\mbox{if }
({\mathrm d}s\wedge {\mathrm d}I)/({\mathrm d}x\wedge {\mathrm d}y)>0 \mbox{ at }(x,y), \\
\sigma (x,y)\equiv -1, & \quad\mbox{if }
({\mathrm d}s\wedge {\mathrm d}I)/({\mathrm d}x\wedge {\mathrm d}y)<0 \mbox{ at }(x,y).
\end{array}
\right.
\]
Here the operator $\wedge $ means outer product.
\par
From the above discussions it is clear that,
\\ \\
{\bf Theorem 4.1} {\it If a symplectic map {\rm (\ref{symplectic2})} has a first integral $I(x,y)$
and the restricted invariant measure {\rm (\ref{InvMeas})} of a generic level set $I=C$ is
finite, then the restricted dynamics $T|_{I=C}$ is isomorphic (mod 0) to an interval exchange,}
\\ \\
where a generic level set was already explained in this section. \\
Moreover, if the function $f(x)$ of (\ref{symplectic2}) is continuous or rational, we can conclude
that,
\\ \\
{\bf Theorem 4.2} {\it In the same assumptions as in theorem {\rm 4.1}, if $f(x)$ is continuous
or rational, then the interval exchange is integrable.}
\\ \\
To prove the theorem 4.2, first we consider the case that $f(x)$ is continuous. In such cases, the
restricted dynamics $T|_{I=C}$ is continuous, too. The case that the level set $I=C$ is connected
is the simplest case of these, and, in this case, $T|_{I=C}$ couldn't have any proper rifts.
Here ``proper rift" is a rift which may not be lost in successive iterations. While the
discrete rotations are the only cases of the interval exchanges which have no proper rifts.
So, the continuous map $T|_{I=C}$ for connected $I=C$ is isomorphic to a discrete rotation,
which is integrable. \\
While if $I=C$ is decomposed to some connected components $M_i$, $i=1,2,\dots $, (we ignore
any measureless connected component, namely, $\mu |_{I=C}(M_i)>0$ for every $i$,)
then the image of
each $M_i$ is also connected because of the continuity of $T|_{I=C}$. Thus, there exists the
indices $j(i)$ for each $i$ such that $M_{j(i)}\supset T|_{I=C}(M_i)$.
The same discussion can be performed for the inverse map $T|_{I=C}^{-1}$ since the continuity of
$f(x)$ makes $T|_{I=C}^{-1}$ continuous, too. Thus, we can get the relation
$T|_{I=C}^{-1}(M_{j(i)})\subset M_i$, which is equivalent to $M_{j(i)}\subset T|_{I=C}(M_i)$.
Together with the above relation, we can get the relation $M_{j(i)}=T|_{I=C}(M_i)$.
Since the restricted invariant measure (\ref{InvMeas}) is finite, there exist finite numbers
$n(i)$ for all indices $i$ such that $T|_{I=C}^{n(i)}(M_i)=M_i$. If we denote $T^{n(i)}|_{M_i}$ as
the $n(i)$ iterations of $T$ restricted on $M_i$, then all these maps are continuous and all these
domains $M_i$ are connected. Thus, the $T^{n(i)}|_{M_i}$ for every $i$ is isomorphic to a discrete
rotation. From the integrability of $T^{n(i)}|_{M_i}$, we can get the conclusion that $T|_{M_i}$
is integrable. Indeed, we can take the decomposition
$T^m|_{M_i}=(T^{n(i)}|_{M_i})^{\alpha }\cdot T^{\beta }|_{M_i}$, where $\alpha $ and $\beta$ are
the quotient and the remainder of $m/n(i)$, respectively, therefore, the maps $T|_{M_i}$ are
solvable by quadratures and finite(, bounded above by $n(i)-1$,) iterations, that is,
are solvable by quadratures.
\par
Second, we consider the other case, namely, the case that $f(x)$ is rational. In such cases,
the continuity of $T|_{I=C}$ breaks down around the first type singular points. While the
neighborhood of every first type singular point returns back to a neighborhood of a finite point
by successive iterations since the invariant measure (\ref{InvMeas}) is finite.
Namely, the difference equations pass the singularity confinement test in some step.
One may think that, for some singular point, the corresponding point which the singular point
{\it returns back} to
and the number of steps for the right limit can be different from for the left limit.
But actually, such bad situations can not occur since $f(x)$ is rational.
Let $p$ be a first type singular point on $I=C$ and $p_+$ and $p_-$ be the left and right limit
of $p$, respectively, and $N_p$ be the number of steps at which neighborhood of $p$ returns back
to a finite region.
Although every element of the sequences $T^n|_{I=C}(p_+)$, $T^n|_{I=C}(p_-)$,
$n=1,2,\dots ,N_p-1$, diverges to infinite point, we can regard that $T^n|_{I=C}(p_+)$ and
$T^n|_{I=C}(p_-)$ for any fixed $n$ are connected to each other, with no paradoxes.
Then the topology of the lavel set $I=C$ is changed to that the map $T|_{I=C}$ is continuous
on $I=C$ for new topology. Therefore, we can perform the same discussion that we performed for
continuous $f(x)$, and thus, we can conclude that the corresponding interval exchanges are
integrable. The proof of theorem 4.2 is finished.
\par
And moreover,
\\ \\
{\bf Corollary 4.3} {\it under the same assumptions as in theorem {\rm 4.2}, the map is onset
linearizable to an elliptic equations.}
\\ \\
An example of difference equations whose integrability and linearizability are
proved by the result stated in this section is presented in appendix.
\section{Discussions}
In this paper, we treat a class of two dimensional difference equations, that is,
symplectic difference equations with one degree of freedom. We get sufficient conditions
that such a difference equation can be reduced to interval exchange (theorem 4.1), and that
such a difference equation is integrable (theorem 4.2). Note that, in this paper, we adopt the
concept that the system can be solved by quadratures as the definition of integrability.
We also partially reveal when singularity confinement test works as the indicator of integrability.
\par
The result is naturally extended to higher degrees as the analogue of complete integrability.
But, how we should extend to higher order as the analogue of Liouville-Arnold's integrability
is not so clear. This is one of the future's assignments.
\par
The advantage of the reduction for a symplectic map to an interval exchange is easiness to treat.
If the map (\ref{symplectic2}) for an irrational $f(x)$ passes to the assumptions in theorem 4.1,
then the $T|_{I=C}$ is isomorphic (mod 0) to an interval exchange, which is expected to have
singular continuous spectrum. To find such an example is also one of the future's assignments.

\section*{Acknowledgement}
The auther thanks Dr. Kazuhiro Nozaki, Dr. Takahiro Morisaki and Dr. Tetsuro Konishi
for useful discussion and helpful advices.
\appendix
\section{Example}
In this section, an example whose integrability is proven by the main theorems is presented.
Its onset linearizability is explained, too.
\par
\begin{figure}
\begin{center}
\includegraphics[height=8cm, width=8cm]{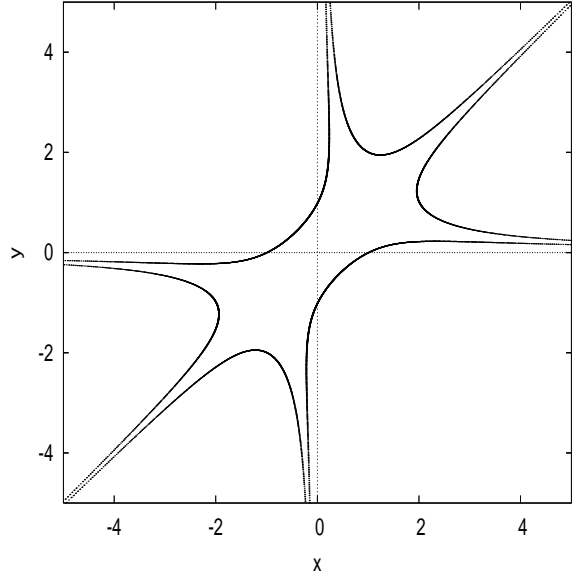}
\caption{An orbit of (\ref{ex1}) on its phase space.}
\label{FigA}
\end{center}
\end{figure}

Consider a system of symplectic difference equations
\begin{equation}
\left\{
\begin{array}{l}
x_{n+1}=y_n \\
y_{n+1}=-x_n+y_n+\frac{1}{y_n}.
\end{array}
\right.
\label{ex1}
\end{equation}
This system has a conserved quantity $I(x,y)=(x-y)^2(xy-1)^2$; see fig.\ref{FigA}.
Therefore, the invariant measure
\begin{equation}
\mu|_{I=C}=\frac{{\rm d}s}{|\mbox{grad}I|}=\frac{-I_y{\rm d}x+I_x{\rm d}y}{|\mbox{grad}I|^2}
=-\frac{{\rm d}x}{I_y}=\frac{{\rm d}y}{I_x}
\label{ex1inv1}
\end{equation}
restricted on $I=C>0$ is finite. Here the third and forth equalities is calculated with the
relation $I_x{\rm d}x+I_y{\rm d}y=0$ on the level set. Thus, the the normalized invariant measure
${\bar \mu }|_{I=C}$ is given as
\begin{equation}
{\bar \mu }|_{I=C}=\frac{1}{6A(C)+8B(C)}\mu|_{I=C},
\label{ex1inv2}
\end{equation}
where, $A(C)$ and $B(C)$ are
\[
A(C)=\int_{-\sqrt{C}}^0\frac{{\rm d}x}{2\sqrt{C}\sqrt{(x^2-1)^2-4\sqrt{C}x}}
\]
and
\[
B(C)=\int_0^{\infty }\frac{{\rm d}x}{2\sqrt{C}\sqrt{(x^2-1)^2-4\sqrt{C}x}},
\]
respectively.
\par
Every level set is composed of six curves
\[
S_1=\{ (x,y)\ |\ I(x,y)=C,\ xy-1<0,\ y-x>0\} ,
\]
\[
S_2=\{ (x,y)\ |\ I(x,y)=C,\ xy-1>0,\ y-x<0,\ x<0\} ,
\]
\[
S_3=\{ (x,y)\ |\ I(x,y)=C,\ xy-1>0,\ y-x<0,\ x>0\} ,
\]
\[
S_4=\{ (x,y)\ |\ I(x,y)=C,\ xy-1<0,\ y-x<0\} ,
\]
\[
S_5=\{ (x,y)\ |\ I(x,y)=C,\ xy-1>0,\ y-x>0,\ x>0\} ,
\]
\[
S_6=\{ (x,y)\ |\ I(x,y)=C,\ xy-1>0,\ y-x>0,\ x<0\}.
\]
Then, we define a map $U$ from the curve $I=C$ to $[0,1)$ as follows:
First, set the image of the tail $(\infty ,0)$ of $S_1$ by $U$ to be $0$, and the image of any
other point $(x,y)\in S_1$ to be $\lambda_1(x,y)$, which is the distance from $(\infty ,0)$ to
$(x,y)$ measured by the metric (\ref{ex1inv2}). Then the head $(0,\infty )$ of $S_1$ is mapped to
$(A+2B)/(6A+8B)$.
Next, set the image of the tail of the next $S_i$, namely $(0,-\infty )$ of $S_2$, to be
$(A+2B)/(6A+8B)$, and the image of any other point $(x,y)\in S_2$ to be
$(A+2B)/(6A+8B)+\lambda_2(x,y)$, where $\lambda_2(x,y)$ is the distance from the tail of $S_2$
to $(x,y)$. And so on. Then, the head of $S_3$, or the tail of $S_4$, is mapped to $1/2$, and the
head of $S_6$ is mapped to $1/2$.
\par
\begin{figure}
\begin{center}
\includegraphics[height=6cm, width=6cm]{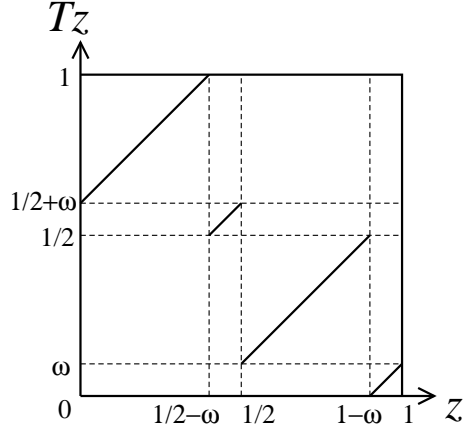}
\caption{The interval exchange map which is isomorphic to onset dynamics of the map (\ref{ex1}).
Here $\omega =\omega (C)$ is defined as $\omega\equiv\frac{A+B}{6A+8B}$}
\label{FigB}
\end{center}
\end{figure}

The interval exchange $T=U\circ\varphi\circ U^{-1}$ is represented at fig.\ref{FigB}.
Here $\varphi :\mbox{\Bb R}^2\to\mbox{\Bb R}^2$ is the map defined by (\ref{ex1}).
Pay your attention to that the interval exchange $T$ depends on $C$,
because $A$ and $B$ depend on $C$.
Therefore, if $A(C)=B(C)$, for example, every point on this level set is periodic point with
period $14$. More generally, if $A(C)$ and $B(C)$ are rationally dependent, every point
on this level set is periodic point with same period, while if the values of $A(C)$ and $B(C)$
are rationally independent, all points on this level set are not periodic.
$(-1,-1)$ and $(1,1)$ are the only isolated periodic (fixed) points of the system (\ref{ex1}).
\par
The interval exchange $T:[0,1)\to [0,1)$ has the property of discrete rotations, that is,
$T^2$ on $[0,1/2)$ or $[1/2,1)$ is a discrete rotation. Its winding number is
$2\omega =(A+B)/(3A+4B)$. Therefore we can solve the dynamics of $T$ as
\[
T^{2n}z=
\left\{
\begin{array}{ll}
z+2\omega n\ \mbox{mod}\ \frac{1}{2} & \quad 0\leq z<\frac{1}{2} \\
\left( z+2\omega n\ \mbox{mod}\ \frac{1}{2}\right)+\frac{1}{2} & \quad\frac{1}{2}\leq z<1,
\end{array}
\right.
\]
\[
T^{2n+1}z=
\left\{
\begin{array}{ll}
\left( z+\omega (2n+1)\ \mbox{mod}\ \frac{1}{2}\right)+\frac{1}{2}
& \quad 0\leq z<\frac{1}{2} \\
z+\omega (2n+1)\ \mbox{mod}\ \frac{1}{2}
& \quad\frac{1}{2}\leq z<1.
\end{array}
\right.
\]
\par
Moreover, this result says that the system (\ref{ex1}) is indirectly onset linearizable.
If we use new coordinate $(\xi_n,\eta_n)=(\cos 4\pi z_n,\sin 4\pi z_n)$, $z_n\equiv T^nz$,
we can describe the onset dynamics as
\[
\left(
\begin{array}{l}
\xi_{n+2} \\
\eta_{n+2}
\end{array}
\right)
=
\left(
\begin{array}{cc}
\cos 4\pi\omega & -\sin 4\pi\omega \\
\sin 4\pi\omega & \cos 4\pi\omega
\end{array}
\right)
\left(
\begin{array}{l}
\xi_n \\
\eta_n
\end{array}
\right) .
\]
As the notations in section 2,
this indirectly onset linearization is carried out with $k=2$, $i=2$.

\end{document}